\documentclass[a4paper,11pt]{article}

\usepackage{graphicx,amssymb}
\begin{document}
\topmargin 0pt \oddsidemargin 0mm
\renewcommand{\thefootnote}{\fnsymbol{footnote}}
\begin{titlepage}
\begin{flushright}
\end{flushright}

\vspace{5mm}
\begin{center}
{\Large \bf  Coincidence Problem in an Oscillating Universe}
\vspace{12mm}

{\large
Guangcan Yang \footnote{Email address: yanggc@wznc.zj.cn }\\
\vspace{8mm} { \em School of Physics and Electronic Information,\\
Wenzhou University, Wenzhou 325000, China}}

\vspace{12mm}

{\large
Anzhong Wang \footnote{Email address: anzhong\_wang@baylor.edu }\\
\vspace{8mm} { \em CASPER, Department of Physics, Baylor
University, \\
Waco, TX 76798-7316, USA}}

\end{center}
\vspace{5mm}
\centerline{{\bf{Abstract}}}
 \vspace{5mm}
 We analyze an oscillating universe model in  brane world
 scenario. The oscillating universe cycles through a series of
 expansions and contractions and its energy density is dominated
 by dust matter at early-time expansion phase and by phantom dark energy
  at late-time expansion phase. We find that the period of the
  oscillating universe is not sensitive to the tension of the
  brane, but sensitive to the equation-of-state parameter $w$ of the
  phantom dark energy, and the ratio of the period to the current
  Hubble age approximately varies from $3$ to $9$ when the
  parameter $w$ changes from $-1.4$ to $-1.1$.  The fraction of
  time that the oscillating universe spends in the coincidence
  state is also comparable to the period of the oscillating universe.
  This result indicates that the coincidence problem can be significantly
  ameliorated in the oscillating universe without singularity.

\end{titlepage}

\newpage
\renewcommand{\thefootnote}{\arabic{footnote}}
\setcounter{footnote}{0} \setcounter{page}{2}

One of remarkable discoveries over the past few years is that
expansion of the universe is speeding up, rather than slowing
down~\cite{dark1,dark2}.  To explain the accelerated expansion, a
so-called dark energy component with large negative pressure is
needed in the energy density of the universe. A lot of evidence
from astronomical observations indicates that our universe is
spatial flat and consists of approximately $73\% $ dark energy,
$27\%$ dust matter including cold dark matter and baryon matter,
and negligible radiation. Although some proposals explaining dark
energy have been suggested, it is fair to say that the dark energy
problem is still a big challenge to theorists and cosmologists,
for recent reviews see \cite{review}

A simple candidate of dark energy is a tiny positive cosmological
constant. In this case, one has to explain why the cosmological
constant is so small, rather than its natural expectation, namely
the Planck energy scale. This is the cosmological constant
problem. Another puzzle associated with dark energy is why the
dark energy density and dust matter energy density are comparable
just now, or why the universe begins its accelerated expansion
just only recently? This is called  cosmic coincidence problem. To
give a solution to the coincidence problem, some models like
quintessence~\cite{quit}, k-essence~\cite{k-essence,k2},
k-chameleon~\cite{k-cham} etc., have been put forward. Of course,
that the energy density of dark energy and that of dust matter are
in the same order now might just be a coincidence, and does not
imply any special meaning.

Suppose that the dark energy has the equation of state, $p=w\rho$,
where $p$ and $\rho$ are the pressure and energy density of dark
energy, respectively. In order for the universe to accelerated
expand, the equation-of-state parameter $w$ has to have $w<-1/3$.
Current observation data give constraint: $-1.46 <w
<-0.78$~\cite{obser}.  The cosmological constant corresponds to a
perfect fluid with  $w=-1$. The quintessence model has $-1 <w
<-1/3$, while the k-essence has $ -1<w<-1/3$ or $w<-1$, but cannot
cross $w=-1$~\cite{vik}. It is well-known that if some matter has
 $w<-1$, it violates all
energy conditions. In that case, some strange things will happen.
The matter with $w<-1$ is called phantom matter, and dark energy
with $w<-1$ is dubbed phantom dark energy~\cite{phantom}. Indeed,
in the phantom dark energy model, a remarkable feature is that the
universe will end  with a further singularity (big rip), where the
scale factor of the universe, energy density of phantom matter
etc. diverge. This implies that the lifetime of a phantom
dominated universe is finite. Thus it is possible to ameliorate
the coincidence problem in a phantom dominated universe if the
fraction of the total lifetime of the universe in a state for
which the dark energy and dark matter densities are roughly
comparable is not so small~\cite{Mic}. Indeed, lately
Scherrer~\cite{Sch} has carried out such a calculation by defining
what means by that dark energy and dark matter energy densities
are roughly comparable. Defining $r=\rho_{e}/\rho_{m}$, where
$\rho_e$ and $\rho_m$ are energy densities of dark energy and dust
matter, respectively, and setting a certain value $r_0$ of $r$, if
$ 1/r_0 < r <r_0$, it is regarded as that the two energy densities
are comparable and the universe is in the coincidence state. For
$r_0=10$, Scherrer found that the fraction varies from $1/3$ to
$1/8$ as $w$ changes from $-1.5$ to $-1.1$; the fraction is
smaller for smaller values of $r_0$. Indeed the coincidence
problem is significantly ameliorated in the sense that the
fraction is not so small. Following \cite{Sch}, Cai and
Wang~\cite{CW} studied the coincidence problem in an interacting
phantom dark energy model with dark matter, and Avelino~\cite{Ave}
discussed the coincidence problem  in a scalar field dark energy
model with a linear effective potential. Note that the phantom
dominated universe is characterized with a further singularity,
while the universe with scalar field dark energy model with a
linear effective potential will collapse to a big crunch. In this
note we will investigate the coincidence problem for a cyclic
universe model in brane world scenario.

 Over the past few years one of most important
progresses in gravity theory is the proposal that our unverse is a
3-brane embedded in a higher dimensional spacetime, the so-called
brane world scenario. In this scenario all standard model matters
are confined on the brane, while gravity can propagate in the
whole spacetime. Among brane world models, the RSII
model~\cite{RSII} is very attractive and is studied intensively in
the literature. In the RSII model, the brane is embedded in a five
dimensional anti-de Sitter space, due to a warped factor, the four
dimensional general relativity on the brane is recovered in the
low energy limit.  And the Friedmann equation for a flat
Friedmann-Robertson-Walker universe on the brane gets modified as
follows~\cite{Fre}
\begin{equation}
\label{eq1}
 H^2= \frac{8\pi G}{3}\rho \left(
1+\frac{\rho}{2\sigma}\right),
\end{equation}
where $H=\dot a/a $ denotes the Hubble parameter, $G$ is the
Newtonian constant on the brane, $\sigma$ is the tension of the
brane and $\rho$ is the energy density of matter on the brane.
Clearly when $\rho \ll \sigma$, the four dimensional general
relativity is recovered. The nucleosynthesis limit gives the
constraint on the tension $\sigma > (1Mev)^4$, while the Newtonian
law of gravity at distance $r\sim 1mm$ imposes the condition
$\sigma
> (10^3 Gev)^4$~\cite{Marrt}. Note that the extra dimension is
spacelike in the RSII model. It is possible that the extra
dimension is timelike~\cite{SS}. In such a model, the Friedmann
equation turns out to be~\cite{SS}
\begin{equation}
\label{eq2}
 H^2= \frac{8\pi G}{3}\rho \left(
1-\frac{\rho}{2\sigma}\right).
\end{equation}
It is remarkable that one can build an oscillating universe model
through phantom dark energy~\cite{cyclic}. At the early-time
expanding phase, the universe is dominated by dust matter.  The
phantom dark energy grows rapidly and dominates the late-time
expanding phase. It can be seen from (\ref{eq2}) that the universe
start expansion at $\rho_{cr}=2\sigma$, where dust matter
dominates the energy density of the universe and stop its
expansion at the critical density $\rho_{cr}= 2\sigma$, where
phantom dark energy dominates the energy density of the universe,
and then start contraction until the energy density reaches to its
critical value again. After this the universe will expand to the
critical value  again. In this way, the universe finishes  a
cycle. Clearly this cosmology is singularity-free. The goal of
this note is  to discuss the cosmological model and the cosmic
coincidence problem for this oscillating universe. Concretely we
will calculate the fraction of a period of the universe when the
universe is in the coincidence state by generalizing the
discussions in \cite{Sch,CW,Ave} to the cyclic universe  model.

Consider a cosmological model, in which all energy components
consist of dust matter (including cold dark matter and  baryon
matter) with $w=0$ and phantom dark energy with a constant
equation-of-state parameter $w<0$.  In this note we will not
consider possible interaction between dust matter and dark energy.
In this case, dust matter has the vanishing equation-of-state
parameter, $w=0$, its energy density satisfies
\begin{equation}
\label{eq3}
\rho_m = \rho_{m0}a_0^3/a^3,
\end{equation}
where $a$ is the scale factor of the universe, $\rho_{m0}$ is a
constant and $a_0$ is the present value of the scale factor. For a
flat universe, one can set $a_0=1$, then $\rho_{m0}$ can be
explained as the current energy density of dust matter. So from
now on we set $a_0=1$.  According to the continuity equation,
\begin{equation}
\dot \rho +3 H(\rho +p)=0,
\end{equation}
the energy density of phantom dark energy has the form
\begin{equation}
\label{eq5}
\rho_e=\rho_{e0}a^{-3(1+w)},
\end{equation}
where  $\rho_{e0}$ is the current dark energy density.
Substituting (\ref{eq3}) and (\ref{eq5}) into (\ref{eq2}), one has
\begin{equation}
\label{eq6}
 H^2 =H_0^2 \left ( \Omega_{m0} a^{-3}
+\Omega_{e0}a^{-3(1+w)}
  \right )
   \frac{1-(\rho_{m0}a^{-3}+\rho_{e0}a^{-3(1+w)}) /2\sigma }{
     1-(\rho_{m0}+\rho_{e0})/2\sigma },
   \end{equation}
  where $H_0$ is the current Hubble parameter, $\Omega_{m0}$ and $\Omega_{e0}$ are
  current fraction
  energy densities of dust matter and dark energy, respectively.
  Integrating (\ref{eq6}) yields
  \begin{equation}
  \label{eq7}
  t =H_0^{-1} \int da \ a ^{1/2}(1-\Omega_{e0}(1-a^{-3w}))^{-1/2}
   (1-\frac{\rho_{m0}a^{-3}
   +\rho_{e0}a^{-3(1+w)}}{2\sigma})^{-1/2},
   \end{equation}
   where we have considered that current energy density of the
   universe is much less than the brane tension, that is,
   $\rho_{m0}+\rho_{e0} \ll \sigma$, and $\Omega_{m0}+\Omega_{e0}
    \approx 1$. Note that the total energy density of the universe
    is given by
\begin{equation}
\rho = \rho_{m0} a^{-3} +\rho_{e0} a^{-3(1+w)}.
\end{equation}
Clearly the energy density is dominated by the dust matter at the
early-time expansion phase, and by the phantom dark energy at
late-time expansion  because $w<-1$. Therefore the two turning
points, satisfying $\rho=2\sigma$, approximately are
\begin{equation}
\label{eq9}
 a_{min} \approx
\left(\frac{\rho_{m0}}{2\sigma}\right)^{1/3},\ \ \ a_{max} \approx
\left(\frac{2\sigma}{\rho_{e0}}\right)^{1/(3|1+w|)}.
\end{equation}
When the oscillating universe  reaches the maximal scale factor
$a_{max}$ from the minimal one $a_{min}$, it takes
\begin{equation}
\label{eq10}
 T =H_0^{-1} \int^{a_{max}}_{a_{min}} da \ a ^{1/2}(\Omega_{m0}+ \Omega_{e0}a^{-3w})^{-1/2}
   (1-\frac{\rho_{m0}a^{-3}
   +\rho_{e0}a^{-3(1+w)}}{2\sigma})^{-1/2}.
   \end{equation}
Denote the current energy density of the universe by
$\rho_0=\rho_{m0}+\rho_{e0}$, one has  $\rho_0/\sigma \ll 1$,
where $\rho_0$ is the current critical density given by
$\rho_0=8\pi G H_0^2/3$. Further defining  $s \equiv
\rho_0/2\sigma$, we can rewrite (\ref{eq9}) and (\ref{eq10}) as
\begin{equation}
a_{min} \approx (s\Omega_{m0})^{1/3},\ \ \ a_{max} \approx
 (s\Omega_{e0})^{-1/(3|1+w|)},
\end{equation}
and
\begin{equation}
\label{eq12}
 T =H_0^{-1} \int^{a_{max}}_{a_{min}} da \ a ^{1/2}(\Omega_{m0}+ \Omega_{e0}a^{-3w})^{-1/2}
   (1-s (\Omega_{m0}a^{-3}
   +\Omega_{e0}a^{-3(1+w)}))^{-1/2},
   \end{equation}
   respectively. According to current observation data, $H_0= 72
   km \cdot s^{-1} Mpc^{-1}$,  which gives the critical energy
   density $\rho_0= 0.42\times 10^{-46}Gev^4$ and current
    Hubble age of the universe  $H_0^{-1}=13.58 Gyr$.
   Note that the $\Lambda$CDM model tells us that the age of the universe
   is about $13.7Gyr$.
    In addition, we
   will take $\Omega_{m0}=0.27$ and $\Omega_{e0}=0.73$ in what
   follows~\cite{WMAP}. In Fig.~1 we plot the ratio of the period of the
   oscillating universe to the current Hubble age $H_0^{-1}$ for
   the case of $\sigma= (10^3Gev)^4$. The ratio approximately varies from
   $3$ to $9$ as $w$ changes from $-1.4$ to $-1.1$.  In particular, we find that
   the ratio is not sensitive to the tension $\sigma$  of the
   brane. For example, if $\sigma=(1Mev)^4$, the ratio for this
   case is almost indistinguishable to the case of
   $\sigma=(10^3Gev)^4$. The insensitivity can be understood as
   follows.

\begin{figure}[ht]
\centering
\includegraphics[totalheight=1.7in]{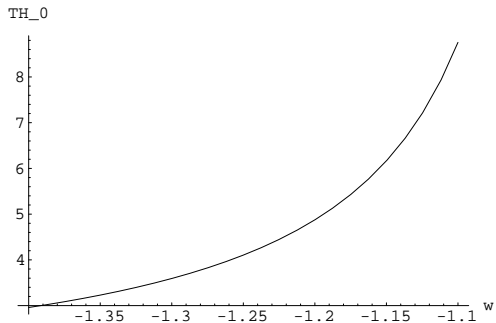}
\caption{The ratio $T/H_0^{-1}$  of the period $T$ of the
oscillating universe to the current Hubble age $H_0^{-1}$ when
$\sigma=(10^3Gev)^4$.}
\end{figure}

The integrand in (\ref{eq12}) diverges at $a_{min}$ and $a_{max}$,
and it decreases very quickly from infinity at $a_{min}$ to an
almost vanishing value at some place. It remains almost unchanged
until $a$ reaches a value near $a_0=1$, where it starts increasing
to a finite value when $a$ gets around $a_0=1$, and then
decreases again until $a$ arrives at a point near $a_{max}$, from
which the integrand increases again very quickly to infinity at
$a_{max}$. We plot a sketch of the integrand in Fig.~2.
\begin{figure}[ht]
\centering
\includegraphics[totalheight=1.7in]{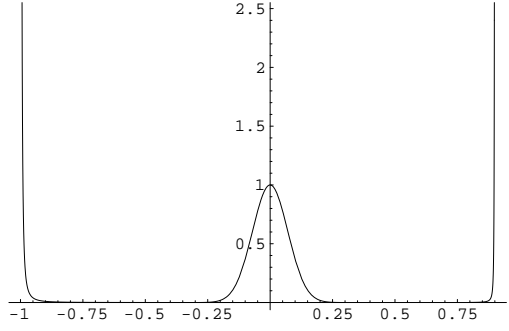}
\caption{A sketch of the integrand in the integration
(\ref{eq12})}
\end{figure}
 We can separate the integration in
(\ref{eq12}) into three parts,
\begin{equation}
T =T_1+T_2+T_3,
\end{equation}
where
\begin{eqnarray}
\label{eq14}
 T_1 &=& H_0^{-1} \int^{a_{1}}_{a_{min}} da \ a
^{1/2}(\Omega_{m0}+ \Omega_{e0}a^{-3w})^{-1/2}
   (1-s (\Omega_{m0}a^{-3}
   +\Omega_{e0}a^{-3(1+w)}))^{-1/2} \nonumber \\
   & \approx &H_0^{-1} \int^{a_{1}}_{a_{min}} da \frac{a ^{1/2}}{
   \sqrt{\Omega_{m0}}\sqrt{
   1-s \Omega_{m0}a^{-3}}}  \nonumber \\
   &=& H_0^{-1} \frac{2}{3\sqrt{\Omega_{m0}}} \sqrt{a^3-\Omega_{mo}s} \left
   |^{a_1}_{a_{min}} \right.
   = H_0^{-1}(1.28 \sqrt{a_1^3-a_{min}^3}),
   \end{eqnarray}
   \begin{equation}
T_2 =H_0^{-1} \int^{a_{2}}_{a_{1}} da \ a ^{1/2}(\Omega_{m0}+
\Omega_{e0}a^{-3w})^{-1/2}
   (1-s (\Omega_{m0}a^{-3}
   +\Omega_{e0}a^{-3(1+w)}))^{-1/2},
   \end{equation}
and
\begin{eqnarray}
\label{eq16}
 T_3 &=& H_0^{-1} \int^{a_{max}}_{a_{2}} da \ a
^{1/2}(\Omega_{m0}+ \Omega_{e0}a^{-3w})^{-1/2}
   (1-s (\Omega_{m0}a^{-3}
   +\Omega_{e0}a^{-3(1+w)}))^{-1/2} \nonumber \\
   &\approx & H_0^{-1} \int^{a_{max}}_{a_{2}} da \ \frac{a
   ^{(1+3w)/2}}{\sqrt{\Omega_{e0}}\sqrt{1-s\Omega_{e0}a^{-3(1+w)}}}
     \nonumber \\
   &=& H_0^{-1} \left( \frac{2 a^{3(1+w)/2}}{3(1+w)\sqrt{\Omega_{e0}}}
 \sqrt{1-s\Omega_{e0}a^{-3(1+w)}} \right)
 \left |^{a_{max}}_{a_{2}} \right. \nonumber \\
 &=& H_0^{-1}\left
 (-0.78\frac{a_2^{3(1+w)/2)}}{1+w}\sqrt{1-\frac{a_2^{-3(1+w)}}{a_{max}^{-3(1+w)}}}
    \right)
   \end{eqnarray}
 respectively. During $T_1$, the scale factor expands from
 $a_{min}$ to $a_1$, and the dust matter is dominant in the energy
 density of the universe, where the phantom dark energy can be
 neglected. On the other hand, during $T_3$ the scale factor
 varies from $a_2$ to $a_{max}$, and the phantom dark energy is
 dominant. During $T_2$, both terms, dust matter and dark energy,
 make contributions. We note that the contributions to $T$ of
 both terms, $T_1$ and $T_3$, are very small, compared to that of
 $T_2$.
 Take an example: if $\sigma= (10^3Gev)^4$ and $w=-1.4$, one has $s=2.1*
 10^{-59}$, $a_{min} =1.78*10^{-20}$, and
 $a_{max}=0.81*10^{49.2}$. Even we take $a_1=10^{-3}$ and
 $a_2=10^4$, (\ref{eq14}) and (\ref{eq16}) then give us $T_1H_0 = 0.128$ and
 $T_3H_0=0.0078$, respectively. Take another extremal example: if
 $\sigma=(1Mev)^4$ and $w=-1.4$, one then has $s=2.1*10^{-35}$,
 $a_{min}=1.78*10^{-12}$ and $a_{max}=0.7*10^{29.2}$. If  taking
$a_1 = 10^{-3}$ and $a_2=10^4$, one still has $T_1H_0=0.128$ and
$T_3H_0= 0.0078$ since $a_1\gg a_{min}$ and $a_2 \ll a_{max}$.
Furthermore, it can be seen from (\ref{eq14}) and (\ref{eq16})
that a smaller $a_1$ leads to a smaller $T_1H_0$ and a larger
$a_2$ gives a smaller $T_3H_0$. Fig.~1 shows that these small
contributions $T_1H_0$ and $T_3H_0$ are indeed negligible.

Further, if one defines the ratio of the dark energy density to
that of dust matter as
\begin{equation}
r \equiv \frac{\rho_{e}}{\rho_{m}}=
\frac{\Omega_{e0}}{\Omega_{m0}}a ^{-3w},
\end{equation}
the integration (\ref{eq7}) can be rewritten as
\begin{equation}
t= -\frac{1}{3w} \frac{\sqrt{\Omega_{m0}}}{H_0\Omega_{e0}}
   \int dr \left(
   \frac{\Omega_{e0}}{r\Omega_{m0}}\right)^{(1+2w)/2w}
     (1+r)^{-1/2} \left( 1-s \Omega_{m0}(1+r)
     \left( \frac{r \Omega_{m0}}{\Omega_{e0}}\right)^{1/w}\right)^{-1/2},
\end{equation}
and the period $T$ in (\ref{eq10}) can be expressed as
\begin{equation}
\label{eq13}
T = -\frac{1}{3w}
\frac{\sqrt{\Omega_{m0}}}{H_0\Omega_{e0}}
   \int^{r_{max}}_{r_{min}} dr \left(
   \frac{\Omega_{e0}}{r\Omega_{m0}}\right)^{(1+2w)/2w}
     (1+r)^{-1/2} \left( 1-s \Omega_{m0}(1+r)
     \left( \frac{r \Omega_{m0}}{\Omega_{e0}}\right)^{1/w}\right)^{-1/2},
\end{equation}
where
\begin{equation}
r_{min} = \frac{\Omega_{e0}}{\Omega_{m0}}\left(s
\Omega_{m0}\right)^{-w}, \ \ r_{max}=
\frac{\Omega_{e0}}{\Omega_{m0}} \left ( s\Omega_{e0}\right)
^{-w/(1+w)}.
\end{equation}
Following \cite{Sch}, we define a scale $r_0$. If $r$ falls in the
range $1/r_0 <r<r_0$, it is regarded that the universe is in the
coincidence state. The duration $t_U$ of the universe in the
coincidence state is then
\begin{equation}
t_U =-\frac{1}{3w} \frac{\sqrt{\Omega_{m0}}}{H_0\Omega_{e0}}
   \int^{r_0}_{1/r_0} dr \left(
   \frac{\Omega_{e0}}{r\Omega_{m0}}\right)^{(1+2w)/2w}
     (1+r)^{-1/2} \left( 1-s \Omega_{m0}(1+r)
     \left( \frac{r
     \Omega_{m0}}{\Omega_{e0}}\right)^{1/w}\right)^{-1/2}.
\end{equation}
Denote by $g$ the ratio of the duration $t_U$ of the universe in
the coincidence state to the period $T$ of the oscillating
universe, namely,
\begin{equation}
\label{eq22}
 g \equiv \frac{t_U}{T}=\frac
     {\int^{r_0}_{1/r_0} dr \left(
   \frac{\Omega_{e0}}{r\Omega_{m0}}\right)^{(1+2w)/2w}
     (1+r)^{-1/2} \left( 1-s \Omega_{m0}(1+r)
     \left( \frac{r
     \Omega_{m0}}{\Omega_{e0}}\right)^{1/w}\right)^{-1/2}}
     { \int^{r_{max}}_{r_{min}} dr \left(
   \frac{\Omega_{e0}}{r\Omega_{m0}}\right)^{(1+2w)/2w}
     (1+r)^{-1/2} \left( 1-s \Omega_{m0}(1+r)
     \left( \frac{r
     \Omega_{m0}}{\Omega_{e0}}\right)^{1/w}\right)^{-1/2}}.
\end{equation}
We plot in Fig.~3 the ratio $g$ for the case with brane tension
$\sigma= (10^3Gev)^4$. Once again, the ratio is not sensitive to
the tension of the brane. The result shown in Fig.~3 is almost the
same as that showed in Fig.~1 of paper \cite{Sch}, where the
results hold for the case of usual Friedman equation in general
relativity. In other words, the latter is for the case $\sigma \to
\infty$ in (\ref{eq1}) or (\ref{eq2}). But note that the two cases
are quite different for the evolution of the universe: in the
current brane world scenario, the universe is an oscillating one
without singularity, while for a phantom dominated universe in
general relativity the universe begins with a big bang and ends
with a big rip.  This insensitivity can be seen from (\ref{eq22})
that in the numerator because $s$ is a very tiny quantity so that
the factor $(1-s\Omega_{m0}(1+r)(r\Omega_{m0}/\Omega_{e0})^{1/w}$
is negligible during the integration range $1/r_0 <r<r_0$. In the
denominator, although the integrand diverges at $r_{min}$ and
$r_{max}$, as we analyzed above, the contributions from these two
points are small and negligible as well.

\begin{figure}[ht]
\centering
\includegraphics[totalheight=1.7in]{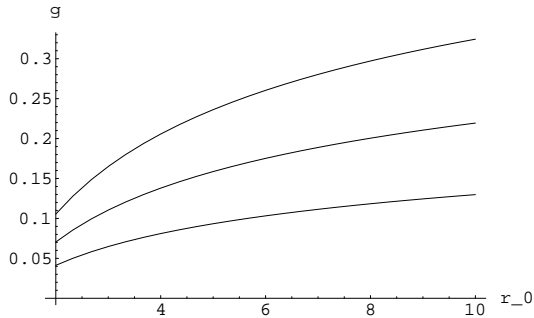}
\caption{The ratio $g$ of the case $\sigma=(10^3Gev)^4$. Three
curves from top to bottom correspond to the cases $w=-1.4$, $-1.2$
and $-1.1$, respectively.}
\end{figure}

Note that for a given $w$ and $r_0$, the fraction of time $g$ that
the oscillating universe spends in the so-called coincidence state
is not small as expected.  We can see from Fig.~3 that for a
larger scale $r_0$, one has a larger $g$; For a fixed scale $r_0$,
a larger $w$ gives a smaller $g$. This can be understood as
follows. From Fig.~1 we note that for a larger $w$, one has a
larger period $T$ of the oscillating universe so that the ratio
$g$ gets smaller. Fig.~3 shows that it is not so strange that we
just live in a period when the dust matter and dark energy
densities are roughly comparable. In this way the coincidence
problem can be significantly ameliorated in this oscillating
universe model without singularity.

In summary we have analyzed an oscillating universe model without
singularity in  brane world scenario. The early-time expansion
phase is dominated by dust matter and  the late-time expansion
phase is dominated by phantom dark energy. We have found that the
period of the oscillating universe is not so sensitive to the
tension of brane, but is sensitive to the equation-of-state
parameter $w$ of phantom dark energy. The ratio of the period to
the current Hubble age of the universe approximately varies from
$3$ to $9$ as $w$ changes from $-1.4$ to $-1.1$. Further we
calculated the fraction of time $g$ that the oscillating universe
spends in the coincidence state. The result shows that it is also
not sensitive to the tension of the brane, but that the fraction
of time of the universe in the coincidence state is comparable to
the period of the oscillating universe.

\section*{Acknowledgments}
The authors thank R. G. Cai for useful suggestions and
discussions.


\begin{thebibliography}{99}
\bibitem{dark1}S.~Perlmutter {\it et al.}  [Supernova Cosmology Project Collaboration],
Astrophys.\ J.\  {\bf 517}, 565 (1999) [arXiv:astro-ph/9812133].

\bibitem{dark2}A.~G.~Riess {\it et al.}  [Supernova Search Team Collaboration],
Astron.\ J.\  {\bf 116}, 1009 (1998) [arXiv:astro-ph/9805201].

\bibitem{review}P.~J.~E.~Peebles and B.~Ratra,
Rev.\ Mod.\ Phys.\  {\bf 75}, 559 (2003) [arXiv:astro-ph/0207347];
T.~Padmanabhan,
Phys.\ Rept.\  {\bf 380}, 235 (2003) [arXiv:hep-th/0212290];
V.~Sahni,
arXiv:astro-ph/0403324.


\bibitem{quit}P.~J.~Steinhardt, L.~M.~Wang and I.~Zlatev,
Phys.\ Rev.\ D {\bf 59}, 123504 (1999) [arXiv:astro-ph/9812313];
I.~Zlatev and P.~J.~Steinhardt,
Phys.\ Lett.\ B {\bf 459}, 570 (1999) [arXiv:astro-ph/9906481].

\bibitem{k-essence}C.~Armendariz-Picon, V.~Mukhanov and P.~J.~Steinhardt,
Phys.\ Rev.\ Lett.\  {\bf 85}, 4438 (2000)
[arXiv:astro-ph/0004134];
C.~Armendariz-Picon, V.~Mukhanov and P.~J.~Steinhardt,
Phys.\ Rev.\ D {\bf 63}, 103510 (2001) [arXiv:astro-ph/0006373].
\bibitem{k2}T.~Chiba, T.~Okabe and M.~Yamaguchi,
Phys.\ Rev.\ D {\bf 62}, 023511 (2000) [arXiv:astro-ph/9912463].

\bibitem{k-cham}H.~Wei and R.~G.~Cai,
arXiv:hep-th/0412045.

\bibitem{obser}R.~A.~Knop {\it et al.},
Astrophys.\ J.\  {\bf 598}, 102 (2003) [arXiv:astro-ph/0309368];
A.~G.~Riess {\it et al.}  [Supernova Search Team Collaboration],
Astrophys.\ J.\  {\bf 607}, 665 (2004) [arXiv:astro-ph/0402512].

\bibitem{vik}A.~Vikman,
Phys.\ Rev.\ D {\bf 71}, 023515 (2005) [arXiv:astro-ph/0407107].

\bibitem{phantom}R.~R.~Caldwell,
Phys.\ Lett.\ B {\bf 545}, 23 (2002) [arXiv:astro-ph/9908168].

\bibitem{Mic}B.~McInnes,
arXiv:astro-ph/0210321.

\bibitem{Sch}R.~J.~Scherrer,
arXiv:astro-ph/0410508.

\bibitem{CW}R.~G.~Cai and A.~Wang,
arXiv:hep-th/0411025.

\bibitem{Ave}P.~P.~Avelino,
arXiv:astro-ph/0411033.

\bibitem{RSII}L.~Randall and R.~Sundrum,
Phys.\ Rev.\ Lett.\  {\bf 83}, 4690 (1999) [arXiv:hep-th/9906064].

\bibitem{Fre}P.~Binetruy, C.~Deffayet and D.~Langlois,
Nucl.\ Phys.\ B {\bf 565}, 269 (2000) [arXiv:hep-th/9905012];
P.~Binetruy, C.~Deffayet, U.~Ellwanger and D.~Langlois,
Phys.\ Lett.\ B {\bf 477}, 285 (2000) [arXiv:hep-th/9910219].

\bibitem{Marrt}R.~Maartens, D.~Wands, B.~A.~Bassett and I.~Heard,
Phys.\ Rev.\ D {\bf 62}, 041301 (2000) [arXiv:hep-ph/9912464].

\bibitem{SS}Y.~Shtanov and V.~Sahni,
Phys.\ Lett.\ B {\bf 557}, 1 (2003) [arXiv:gr-qc/0208047].

\bibitem{cyclic}M.~G.~Brown, K.~Freese and W.~H.~Kinney,
arXiv:astro-ph/0405353.

\bibitem{WMAP}C.~L.~Bennett {\it et al.},
Astrophys.\ J.\ Suppl.\  {\bf 148}, 1 (2003)
[arXiv:astro-ph/0302207];
D.~N.~Spergel {\it et al.}  [WMAP Collaboration],
Astrophys.\ J.\ Suppl.\  {\bf 148}, 175 (2003)
[arXiv:astro-ph/0302209].



\end{thebibliography}
\end{document}